\documentclass[10pt,a4paper]{article}
\usepackage{graphics,tikz,amsmath}

\newcommand{\bea}{\begin{eqnarray}}
\newcommand{\eea}{\end{eqnarray}}
\newcommand{\be}{\begin{equation}}
\newcommand{\ee}{\end{equation}}

\def\be{\begin{eqnarray}}
\def\ee{\end{eqnarray}}
\def\bd{\begin{displaymath}}
\def\ed{\end{displaymath}}

\def\ga{\gamma}
\def\etal{{\em et al}}
\def\ADNDT{{\em At. Data Nucl. Data. Tables }}

\def\NP{{\em Nucl. Phys. }}
\def\PR{{\em Phys. Rev. }}
\def\PRL{{\em Phys. Rev. Lett. }}

\def\EPJ{{\em Eur. Phys. J. }}

\begin{document}
\title{Endpoint of $rp$ process using relativistic mean field approach and a new mass formula}
\author{Chirashree Lahiri$^1$ and G. Gangopadhyay$^2$\\
Department of Physics, University of Calcutta\\
92, Acharya Prafulla Chandra Road, Kolkata-700 009, India\\
email: $^1$chirashree.lahiri@gmail.com, $^2$ggphy@caluniv.ac.in}
\maketitle

\begin{abstract}
Densities from relativistic mean field calculations are applied to 
construct the optical potential and, hence calculate the endpoint of the rapid proton capture ($rp$) process. 
Mass values are taken from a new phenomenological mass formula. 
 Endpoints are calculated for different temperature-density 
profiles of various X-ray bursters. We find that the $rp$ process can  
produce significant quantities of nuclei upto around mass 95.
Our results differ from existing works to some extent.
\end{abstract}

\section{Introduction}

Proton capture reactions at a very low temperature play an important 
role in nucleosynthesis process. Most importantly, in explosive 
nucleosynthesis ({\it e.g.} an X-ray burst), the rapid proton capture ($rp$) process
is responsible for the production of proton-rich isotopes upto mass 100 region.
In nature, the proton capture reactions, important for nucleosynthesis, usually 
involve certain nuclei as targets which are not available on earth
or cannot be produced  in terrestrial laboratories with our present day 
technology. Therefore, theory remains  the sole guide to extract the 
physics. 

In our present work, we have studied the endpoint of the $rp$ process in a 
microscopic approach using a new phenomenological mass formula \cite{mass}. 
In a similar work, Schatz {\em et al.} \cite{schatz} calculated the 
endpoint of $rp$ process  using mass values from a Finite Range droplet Model 
(FRDM)\cite{Mol} 
calculation and  proton capture rates from Hauser-Feshbach code 
NON-SMOKER\cite{ns,ns1}. We will show that the results of our calculation are
different from their observations. In the present work, we will 
concentrate only on X-ray burst scenarios, which have typical timescale of 
100 seconds and a peak proton flux density  of the order of $10^{6} gram/cm^3$. 
This type of burst provides  a highly proton-rich environment  around the peak 
temperatures 1-2 GK. We try to look at different models of the X-ray
burster and find out the endpoint of the $rp$ process nucleosynthesis.

\section{Methodology}

When an X-ray burst takes place, a proton-rich high temperature 
environment, which triggers the $rp$ process, is created. The
process passes through nuclei near the proton drip line, 
not available on earth. In regions far from the stability 
valley, rates derived from phenomenological calculations may not 
represent the reality very well, leading to considerable uncertainty 
in the process. Very often, the reaction rates are varied by a large factor to 
study their effects. On the other hand, in a microscopic calculation, 
uncertainty in reaction rates can be reduced and therefore, this approach 
is expected to give a more accurate result for unknown mass regions. In a previous work\cite{prc},
we have shown that the rates may vary at most by a factor less than two when the cross-section
values range over four orders of magnitude.
 
A microscopic calculation has been performed to evaluate proton 
capture rates for the nuclei involve in the $rp$ process in the present work.
We use the spherical optical model to calculate the rates of the relevant 
reactions. As most of the nuclei involved in the process lie around the drip 
line, experimental density information are not available. Hence, theoretical 
density profiles have been calculated using relativistic mean 
field (RMF) theory. In the present work, we use  the FSU Gold Lagrangian 
density\cite{gold} and
solve the RMF equations in the spherical approximation for the nuclei 
involved in the $rp$ process. This 
Lagrangian density, containing additional nonlinear terms for the vector 
isoscalar meson self interaction and an interaction between the isoscalar vector 
and the isovector vector mesons, has been found to be very useful in describing 
nuclear properties throughout the mass table [See {\em e.g.} Bhattacharya 
\etal\cite{1,2} and references therein].

The microscopic optical model potential for capture
reactions are obtained using effective interactions derived from the nuclear 
matter calculation in local density approximation, {\em i.e.} by 
substituting the nuclear matter density with the density distribution 
of the finite nucleus.  In the present work, we have constructed the 
potential by folding the density dependent M3Y (DDM3Y)\cite{ddm3y1,ddm3y2,ddm3y3} 
interaction with densities from RMF approach. This interaction was extracted 
from a finite-range energy-independent $G$-matrix element of the Reid potential by 
adding a zero-range energy-dependent pseudo-potential and 
a density-dependent factor.
The interaction at the point $\vec{r}$ is, thus, given by
\begin{equation}
 v(r,\rho,E)=t^{M3Y}(r,E)g(\rho)                 
\end{equation} 
where $E$ is the incident energy and $\rho$, the nuclear density. 
The $t^{M3Y}$ interaction is given by
\begin{equation}
t^{M3Y}=7999\frac{e^{-4r}}{4r}-2134\frac{e^{-2.5r}}{2.5r}+J_{00}(E)\delta(r)
\end{equation} 
for $r$ in $fm$, and  $J_{00}(E)$ is the zero range pseudo potential, 
\begin{equation}
J_{00}(E)=-276\left( 1-0.005\frac{E}{A}\right) {\rm MeV} fm^{3}\end{equation} 

The density dependent factor $g(\rho)$ has been chosen of the form 
$C(1-\beta\rho^{2/3})$ from the work by Chaudhuri\cite{ddm3y2,ddm3y3} 
where the constants were obtained
from a nuclear matter calculation as $C$ = 2.07 and
$\beta$ = 1.624 $fm^2$. We have used this form in our calculation
without changing any of the above parameters.

We have also fixed the various parameters and prescriptions in the 
Hauser-Feshbach calculation for the relevant mass region by comparing our
results to the experimental low energy proton capture cross sections for
these nuclei. Our method of calculation for mean field and proton 
capture rates has been described in our earlier works\cite{prc,epja} in detail.
The computer code TALYS\cite{talys} has been used for rate calculation.

Binding energy of nuclear ground state is one of the most important inputs 
in the study of astrophysical reactions. Experimental measurements are very 
difficult to perform in nuclei far from the stability line. 
Therefore, one has to take recourse to theoretical predictions. 
Though we have used a mean field calculation to extract the nuclear density 
profiles, no RMF calculation has been able to achieve a prediction of mass 
values with accuracy sufficient for studying the proton drip line. In fact, 
even Skyrme Hartree-Fock calculations can predict the mass values with 
an root mean square (rms) error slightly less than 0.6 MeV only. Thus, in 
the present work, 
we have obtained the mass values from a newly developed mass 
formula\cite{mass}. It uses a purely phenomenological form with empirically 
fitted parameters and predicts the known mass values of 2140 nuclei with an 
rms error of 0.376 MeV. In view of the success of the formula to predict the proton 
dripline and $rp$ process upto mass 80 region\cite{ijmpe} and to predict the 
peaks in $r$  process\cite{nrich} quite well, it will be interesting 
to see the effect of this mass formula on $rp$ process beyond mass 80 region and 
to the endpoint of the $rp$ process.

In an X-ray burst environment, a nucleus ($Z,A$) may capture a proton to form the 
nucleus ($Z+1,A+1$) . However, this process has to compete with its inverse, 
{\em i.e.} photodisintegration by emitting a proton at high temperature 
[$(\gamma, p)$ reaction]. A negative or a small positive value of the proton 
separation 
energy implies that the inverse reaction dominates and the $rp$ process stalls 
at that point, the so-called waiting point. Therefore, only a two proton
capture process can bridge the waiting point nuclei. The bridging mechanism has been 
discussed in standard text books (for example the book by 
Illiadis\cite{iliadis}). For the $X(p,\gamma)Y$ reaction, the rate $\lambda$ 
of the inverse process $(\gamma,p)$ is 
related to the the proton capture rate by the reciprocity theorem
and is of the form \cite{iliadis} 
 \begin{eqnarray}
\lambda=
9.86851\times10^9 T^{\frac{3}{2}}\left(\frac{M_{p}M_{X}}{M_{Y}}\right)^{\frac{3}{2}}
\frac{\left( 2J_{p}+1 \right)\left( 2J_{X}+1 \right)}
{\left( 2J_{Y}+1 \right)}\frac{G_pG_X}{G_Y}
\nonumber\\
{N\langle\sigma v\rangle_{pX\rightarrow Y\ga}} 
\exp\left(\frac{-11.605Q}{T}\right)
\end{eqnarray}
in the unit of $sec^{-1}$ for  $X(p,\gamma)Y$ process. The forward reaction rate, denoted by 
$N\langle\sigma v\rangle_{pX\rightarrow Y\ga}$, is expressed in 
$cm^3mol^{-1}sec^{-1}$. The 
temperature $T$ is in GK ($10^9$K) and $Q$ is the ground state Q-value of the 
$(p,\gamma)$ reaction  expressed in MeV. The normalized partition functions, 
$G_X$ and $G_Y$, have been obtained from Rauscher \etal.\cite{ptable} For 
protons, we use the standard values,  $G_p=1$ and $J_p=1/2$. 
As evident from the above expression, the Q-value, appearing in the exponent, plays a vital role in the 
whole process.

 Apart from the above processes, a nucleus can decay by emitting beta particles while, for higher 
mass isotopes, $\alpha$-decay is another probable channel. In this work, the measured half life 
values for $\beta$-decay have been taken from the compilation by Audi {\em et al.}\cite{audi14} 
except in the case of $^{65}$As, which is taken from the experimental measurements by L\'{o}pez {
\em et al.}\cite{lopez}  In absence of experimental data, half life values  
have been taken from the calculation by M\"{o}ller {\em et al.}\cite{Mol} both for $\beta$- and 
$\alpha$- decay. Taking into account all the above processes, we have constructed a network 
extended upto A=110 region to study the nucleosynthesis path and relative abundances 
of the elements at any instant.

\section{Results}

 \begin{figure}[b]
\center
\resizebox{10cm}{!}{

\begin{tikzpicture}

  \node (1) at (.4,0)  [draw,inner sep=5.8pt,label=left:$_{28}$Ni,label=below :28] {$ $  };
    \node (2) at (.4,.4) [draw,inner sep=5.8pt] {$ $};
    \node (3) at (.4,.8) [draw,inner sep=5.8pt,label=left:$_{30}$Zn] {$ $};
    \node (4) at (.4,1.2) [draw,inner sep=5.8pt] {$ $};
    \node (5) at (.8,.4) [draw,inner sep=5.8pt]{$ $};
    \node (6) at (.8,.8) [draw,inner sep=5.8pt]{$ $};
 \node (7) at (.8,1.2) [draw,inner sep=5.8pt]   {$ $ };
 \node (8) at (1.2,.4) [draw,inner sep=5.8pt]    {$ $};
 \node (9) at (1.2,.8) [draw,inner sep=5.8pt]    {$ $};
 \node (10) at (1.2,1.2) [draw,inner sep=5.8pt]   {$ $};
\node (11) at (1.2,1.6) [draw,inner sep=5.8pt,label=left:$_{32}$Ge]    {$ $};
 \node (12) at (1.6,.4) [draw,inner sep=5.8pt] {$ $};
 \node (13) at (1.6,.8) [draw,inner sep=5.8pt] {$ $};
 \node (14) at (1.6,1.2) [draw,inner sep=5.8pt] {$ $};
\node (15) at (1.6,1.6) [draw,inner sep=5.8pt]  {$ $};
 \node (16) at (2.0,1.2) [draw,inner sep=5.8pt]   {$ $};
\node (17) at (2.0,1.6) [fill,draw,inner sep=5.8pt]    {$ $};
 \node (18) at (2.0,2.0) [draw,inner sep=5.8pt]   {$ $};
\node (19) at (2.0,2.4) [draw,inner sep=5.8pt]    {$ $};
 \node (20) at (2.4,1.2) [draw,inner sep=5.8pt] {$ $};
 \node (21) at (.8,0) [draw,inner sep=5.8pt]  {$ $};
 \node (22) at (1.2,0) [draw,inner sep=5.8pt,label=below :30]   {$ $};
 \node (23) at (1.6,0) [draw,inner sep=5.8pt] {$ $};
 \node (24) at (2.0,0) [draw,inner sep=5.8pt,label=below :32]   {$ $};
 \node (25) at (2.0,.4) [draw,inner sep=5.8pt]   {$ $};
 \node (26) at (2.0,.8) [draw,inner sep=5.8pt]   {$ $};
\node (27) at (1.2,2.0) [draw,inner sep=5.8pt] {$ $};
\node (28) at (1.2,2.4) [draw,inner sep=5.8pt,label=left:$_{34}$Se] {$ $};
\node (29) at (1.6,2.0) [draw,inner sep=5.8pt] {$ $};
\node (30) at (1.6,2.4) [draw,inner sep=5.8pt] {$ $};
\node (31) at (2.4,1.6) [draw,inner sep=5.8pt] {$ $};
\node (32) at (2.4,2.0) [draw,inner sep=5.8pt] {$ $};
\node (33) at (2.4,2.4) [draw,inner sep=5.8pt] {$ $};
\node (34) at (2.4,2.8) [draw,inner sep=5.8pt] {$ $};
 \node (35) at (2.8,.8) [draw,inner sep=5.8pt,label=below :34] {$ $};
 \node (36) at (2.8,1.2) [draw,inner sep=5.8pt] {$ $};
 \node (37) at (2.8,1.6) [draw,inner sep=5.8pt] {$ $};
\node (38) at (2.8,2.0) [draw,inner sep=5.8pt] {$ $};
\node (39) at (2.8,2.4) [fill,draw,inner sep=5.8pt] {$ $};
\node (40) at (2.8,2.8) [draw,inner sep=5.8pt] {$ $};
\node (41) at (2.8,3.2) [draw,inner sep=5.8pt,label=left:$_{36}$Kr] {$ $};
\node (42) at (3.2,2.0) [draw,inner sep=5.8pt] {$ $};
\node (43) at (3.2,2.4) [draw,inner sep=5.8pt] {$ $};
\node (44) at (3.2,2.8) [draw,inner sep=5.8pt] {$ $};
\node (45) at (3.2,3.2) [draw,inner sep=5.8pt] {$ $};
\node (46) at (3.6,2.4) [draw,inner sep=5.8pt,label=below:36] {$ $ };
\node (47) at (3.6,2.8) [draw,inner sep=5.8pt] {$ $ };
\node (48) at (3.6,3.2) [fill,draw,inner sep=5.8pt] {$ $};
\node (49) at (3.6,3.6) [draw,inner sep=5.8pt] {$ $ };
\node (50) at (3.6,4.) [draw,inner sep=5.8pt,label=left:$_{38}$Sr] {$ $ };
\node (51) at (4.0,2.8) [draw,inner sep=5.8pt] {$ $ };
\node (52) at (4.0,3.2) [draw,inner sep=5.8pt] {$ $  };
\node (53) at (4.0,3.6) [draw,inner sep=5.8pt] {$ $ };
\node (54) at (4.0,4.) [draw,inner sep=5.8pt] {$   $ };
\node (55) at (4.0,4.4) [draw,inner sep=5.8pt] {$  $ };
\node (56) at (4.4,3.2) [draw,inner sep=5.8pt,label=below:38] {$ $ };
\node (57) at (4.4,3.6) [draw,inner sep=5.8pt] {$  $ };
\node (58) at (4.4,4.) [fill,draw,inner sep=5.8pt] {$ $ };
\node (59) at (4.4,4.4) [draw,inner sep=5.8pt] {$ $ };
\node (60) at (4.4,4.8) [draw,inner sep=5.8pt,label=left:$_{40}$Zr] {$ $ };
\node (61) at (4.8,3.6) [draw,inner sep=5.8pt] {$ $ };
\node (62) at (4.8,4.) [draw,inner sep=5.8pt] {$ $ };
\node (63) at (4.8,4.4) [draw,inner sep=5.8pt] {$ $ };
\node (64) at (4.8,4.8) [draw,inner sep=5.8pt] {$ $ };
\node (65) at (4.8,5.2) [draw,inner sep=5.8pt] {$ $ };
\node (66) at (5.2,4.) [draw,inner sep=5.8pt,label=below:40] {$ $ };
\node (67) at (5.2,4.4) [draw,inner sep=5.8pt] {$ $ };
\node (68) at (5.2,4.8) [fill,draw,inner sep=5.8pt] {$ $ };
\node (69) at (5.2,5.2) [draw,inner sep=5.8pt] {$ $ };
\node (70) at (5.2,5.6) [draw,inner sep=5.8pt,label=left:$_{42}$Mo] {$ $ };
\node (71) at (5.6,4.4) [draw,inner sep=5.8pt] {$ $ };
\node (72) at (5.6,4.8) [draw,inner sep=5.8pt] {$ $ };
\node (73) at (5.6,5.2) [draw,inner sep=5.8pt] {$ $ };
\node (74) at (5.6,5.6) [draw,inner sep=5.8pt] {$ $ };
\node (75) at (5.6,6.0) [draw,inner sep=5.8pt] {$ $ };
\node (76) at (6.0,4.8) [draw,inner sep=5.8pt,label=below:42] {$ $ };
\node (77) at (6.0,5.2) [draw,inner sep=5.8pt] {$ $ };
\node (78) at (6.0,5.6) [fill,draw,inner sep=5.8pt] {$ $ };
\node (79) at (6.0,6.0) [draw,inner sep=5.8pt] {$ $ };
\node (80) at (6.0,6.4) [draw,inner sep=5.8pt,label=left:$_{44}$Ru] {$ $ };
\node (81) at (6.4,5.2) [draw,inner sep=5.8pt] {$ $ };
\node (82) at (6.4,5.6) [draw,inner sep=5.8pt] {$ $ };
\node (83) at (6.4,6.0) [draw,inner sep=5.8pt] {$ $ };
\node (84) at (6.4,6.4) [draw,inner sep=5.8pt] {$ $ };
\node (85) at (6.4,6.8) [draw,inner sep=5.8pt] {$ $ };
\node (86) at (6.8,5.6) [draw,inner sep=5.8pt,label=below:44] {$ $ };
\node (87) at (6.8,6.0) [draw,inner sep=5.8pt] {$ $ };
\node (88) at (6.8,6.4) [fill,draw,inner sep=5.8pt] {$ $ };
\node (89) at (6.8,6.8) [draw,inner sep=5.8pt] {$ $ };
\node (90) at (6.8,7.2) [draw,inner sep=5.8pt,label=left:$_{46}$Pd] {$ $ };
\node (91) at (7.2,6.0) [draw,inner sep=5.8pt] {$ $ };
\node (92) at (7.2,6.4) [draw,inner sep=5.8pt] {$ $ };
\node (93) at (7.2,6.8) [draw,inner sep=5.8pt] {$ $ };
\node (94) at (7.2,7.2) [draw,inner sep=5.8pt] {$ $ };
\node (95) at (7.2,7.6) [draw,inner sep=5.8pt] {$ $ };
\node (96) at (7.6,6.4) [draw,inner sep=5.8pt,label=below:46] {$ $ };
\node (97) at (7.6,6.8) [draw,inner sep=5.8pt] {$ $ };
\node (98) at (7.6,7.2) [fill,draw,inner sep=5.8pt] {$ $ };
\node (99) at (7.6,7.6) [draw,inner sep=5.8pt] {$ $ };
\node (100) at (7.6,8.0) [draw,inner sep=5.8pt,label=left:$_{48}$Cd] {$ $ };
\node (101) at (8.0,6.8) [draw,inner sep=5.8pt] {$ $ };
\node (102) at (8.0,7.2) [draw,inner sep=5.8pt] {$ $ };
\node (103) at (8.0,7.6) [draw,inner sep=5.8pt] {$ $ };
\node (104) at (8.0,8.0) [draw,inner sep=5.8pt] {$ $ };
\node (105) at (8.0,8.4) [draw,inner sep=5.8pt] {$ $ };
\node (106) at (8.4,7.2) [draw,inner sep=5.8pt,label=below:48] {$ $ };
\node (107) at (8.4,7.6) [draw,inner sep=5.8pt] {$ $ };
\node (108) at (8.4,8.0) [fill,draw,inner sep=5.8pt] {$ $ };
\node (109) at (8.4,8.4) [draw,inner sep=5.8pt] {$ $ };
\node (110) at (8.4,8.8) [draw,inner sep=5.8pt,label=left:$_{50}$Sn] {$ $ };
\node (111) at (8.8,7.6) [draw,inner sep=5.8pt] {$ $ };
\node (112) at (8.8,8.0) [draw,inner sep=5.8pt] {$ $ };
\node (113) at (8.8,8.4) [draw,inner sep=5.8pt] {$ $ };
\node (114) at (8.8,8.8) [draw,inner sep=5.8pt] {$ $ };
\node (115) at (8.8,9.2) [draw,inner sep=5.8pt] {$ $ };
\node (116) at (9.2,8.0) [draw,inner sep=5.8pt,label=below:50] {$ $ };
\node (117) at (9.2,8.4) [draw,inner sep=5.8pt] {$ $ };
\node (118) at (9.2,8.8) [fill,draw,inner sep=5.8pt] {$ $ };
\node (119) at (9.2,9.2) [draw,inner sep=5.8pt] {$ $ };
\node (120) at (9.2,9.6) [draw,inner sep=5.8pt,label=left:$_{52}$Te] {$ $ };
\node (121) at (9.6,8.) [draw,inner sep=5.8pt] {$ $ };
\node (122) at (9.6,8.4) [draw,inner sep=5.8pt] {$ $ };
\node (123) at (9.6,8.8) [draw,inner sep=5.8pt] {$ $ };
\node (124) at (9.6,9.2) [draw,inner sep=5.8pt] {$ $ };
\node (125) at (9.6,9.6) [draw,inner sep=5.8pt] {$ $ };
\node (126) at (9.6,10.0) [draw,inner sep=5.8pt] {$ $ };
\node (127) at (10.0,8.0) [draw,inner sep=5.8pt,label=below:52] {$ $ };
\node (128) at (10.0,8.4) [draw,inner sep=5.8pt] {$ $ };
\node (129) at (10.0,8.8) [draw,inner sep=5.8pt] {$ $ };
\node (130) at (10.0,9.2) [draw,inner sep=5.8pt] {$ $ };
\node (131) at (10.0,9.6) [fill,draw,inner sep=5.8pt] {$ $ };
\node (132) at (10.0,10.0) [draw,inner sep=5.8pt] {$ $ };
\node (133) at (10.4,8.4) [draw,inner sep=5.8pt] {$ $ };
\node (134) at (10.4,8.8) [draw,inner sep=5.8pt] {$ $ };
\node (135) at (10.4,9.2) [draw,inner sep=5.8pt] {$ $ };
\node (136) at (10.4,9.6) [draw,inner sep=5.8pt] {$ $ };
\node (137) at (10.4,10.0) [draw,inner sep=5.8pt] {$ $ };
\node (138) at (10.8,8.4) [draw,inner sep=5.8pt] {$ $ };
\node (139) at (10.8,8.8) [draw,inner sep=5.8pt] {$ $ };
\node (140) at (10.8,9.2) [draw,inner sep=5.8pt] {$ $ };
\node (141) at (10.8,9.6) [draw,inner sep=5.8pt] {$ $ };
\node (142) at (10.8,10.0) [draw,inner sep=5.8pt] {$ $ };
\node (143) at (11.2,8.4) [draw,inner sep=5.8pt] {$ $ };
\node (144) at (11.2,8.8) [draw,inner sep=5.8pt] {$ $ };
\node (145) at (11.2,9.2) [draw,inner sep=5.8pt] {$ $ };
\node (146) at (11.2,9.6) [draw,inner sep=5.8pt] {$ $ };
\node (147) at (11.6,8.4) [draw,inner sep=5.8pt] {$ $ };
\node (148) at (11.6,8.8) [draw,inner sep=5.8pt] {$ $ };
\node (149) at (11.6,9.2) [draw,inner sep=5.8pt] {$ $ };
\node (150) at (12.0,8.4) [draw,inner sep=5.8pt] {$ $ };
\node (151) at (2.4,.8) [draw,inner sep=5.8pt] {$ $ };
\node (152) at (12.,8.8) [draw,inner sep=5.8pt] {$ $ };
\node (153) at (12.,9.2) [draw,inner sep=5.8pt] {$ $ };
\node (154) at (11.6,9.6) [draw,inner sep=5.8pt] {$ $ };
\node (155) at (12.,9.6) [draw,inner sep=5.8pt] {$ $ };
\node (156) at (12.,10.0) [draw,inner sep=5.8pt] {$ $ };
\node (157) at (11.6,10.0) [draw,inner sep=5.8pt] {$ $ };
\node (158) at (11.2,10.0) [draw,inner sep=5.8pt] {$ $ };

\draw [ thick]  (.4,0) to (.4,.4); 
\draw [very thick]  (.4,.4) to (.4,.8); 
\draw [very thick]  (.4,.8) to (0.8,.4); 
\draw [very thick]  (.8,.4) to (.8,.8); 
\draw [very thick]  (.8,.8) to (1.2,.4); 
\draw [very thick]  (1.2,.4) to (1.2,.8); 
\draw [very thick]  (1.2,.8) to (1.2,1.2); 
\draw [very thick]  (1.2,1.2) to (1.2,1.6); 
\draw [very thick]  (1.2,1.6) to (1.6,1.2);
\draw [very thick]  (1.6,1.2) to (1.6,1.6);
\draw [very thick]  (1.6,1.6) to (2.,1.2); 
\draw [very thick]  (2.,1.2) to (2.,1.6);
\draw [very thick]  (2.,1.6) to (2.,2.);
\draw [gray,thick]  (2.,1.6) to (2.4,1.2);
\draw [gray,thick]  (2.4,1.2) to (2.4,2.0);
\draw [very thick]  (2.,2.) to (2.,2.4);
\draw [very thick]  (2.,2.4) to (2.4,2.);
\draw [very thick]  (2.4,2.) to (2.4,2.4);
\draw [very thick]  (2.4,2.4) to (2.8,2.0);
\draw [very thick]  (2.8,2.) to (2.8,2.4);
\draw [very thick]  (3.2,2.8) to (3.2,3.2);
\draw [very thick]  (3.2,3.2) to (3.6,2.8);
\draw [very thick]  (3.6,2.8) to (3.6,3.2);
\draw [very thick]  (3.6,3.2) to (3.6,3.6);
\draw [very thick]  (3.6,3.6) to (3.6,4.);
\draw [very thick]  (3.6,4) to (4,3.6);
\draw [very thick]  (4,3.6) to (4,4);
\draw [very thick]  (4,4.) to (4.4,3.6);
\draw [very thick] (4.4,3.6) to (4.4,4);
\draw [thick,gray] (4.4,4) to (4.4,4.4);
\draw [thick,gray] (4.4,4.4) to (4.4,4.8);
\draw [thick,gray] (4.4,4.8) to (4.8,4.4);
\draw [very thick] (4.8,4.4) to (4.8,4.8);
\draw [very thick] (4.8,4.8) to (5.2,4.4);
\draw [very thick] (5.2,4.4) to (5.2,4.8);
\draw [very thick] (3.6,3.2) to (4.0,2.8);
\draw [very thick] (4,2.8) to (4,3.2);
\draw [very thick] (4,3.2) to (4,3.6);
\draw [very thick] (4,3.6) to (4,4.);
\draw [very thick] (4.4,4) to (4.8,3.6);
\draw [very thick] (4.8,3.6) to (4.8,4);
\draw [very thick] (4.8,4) to (4.8,4.4);
\draw [very thick] (4.8,4.4) to (4.8,4.8);
\draw [very thick] (2.8,2.4) to (3.2,2.);
\draw [very thick] (3.2,2.) to (3.2,2.4);
\draw [very thick] (3.2,2.4) to (3.2,2.8);
\draw [very thick] (5.2,4.8) to (5.6,4.4);

\draw [very thick] (5.6,4.4) to (5.6,4.8);
\draw [very thick] (5.6,4.8) to (5.6,5.2);
\draw [very thick] (5.6,5.2) to (5.6,5.6);
\draw [very thick] (5.6,5.2) to (6.,4.8);
\draw [very thick] (5.6,5.6) to (6,5.2);
\draw [gray, thick] (6.,4.8) to (6.,5.2);
\draw [very thick] (6.,5.2) to (6.,5.6);
\draw [very thick] (6.,5.6) to (6.4,5.2);
\draw [very thick] (6.4,5.2) to (6.4,5.6);
\draw [very thick] (6.4,5.6) to (6.4,6);
\draw [very thick] (6.4,6) to (6.4,6.4);
\draw [very thick] (6.4,6) to (6.8,5.6);
\draw [very thick] (6.4,6.4) to (6.8,6.);
\draw [gray, thick] (6.8,5.6) to (6.8,6);
\draw [very thick] (6.8,6) to (6.8,6.4);
\draw [very thick] (6.8,6.4) to (7.2,6);
\draw [very thick] (7.2,6) to (7.2,6.4);
\draw [very thick] (7.2,6.4) to(7.2,6.8);
\draw [very thick](7.2,6.8) to (7.2,7.2);
\draw [very thick](7.2,6.8) to (7.6,6.4);
\draw [very thick] (7.2,7.2) to (7.6,6.8);
\draw [very thick] (7.6,6.8) to (7.6,7.2);
\draw [very thick] (7.6,7.2) to (8,6.8);
\draw [very thick] (8,6.8) to (8,7.2);
\draw [very thick] (8,7.2) to (8,7.6);
\draw [very thick] (8,7.6) to (8,8.);
\draw [very thick] (8,7.6) to (8.4,7.2);
\draw [very thick] (8.0,8.0) to (8.4,7.6);
\draw [very thick] (8.4,7.2) to (8.4,7.6);
\draw [very thick] (8.4,7.6) to (8.4,8.0);
\draw [very thick](8.4,8.0) to (8.8,7.6);
\draw [very thick] (8.8,7.6) to (8.8,8.0);
\draw [very thick] (8.8,8.0) to (8.8,8.4);
\draw [very thick] (8.8,8.4) to (9.2,8.0);
\draw [very thick] (9.2,8.0) to (9.2,8.4);
\draw [very thick] (9.2,8.4) to (9.6,8.0);
\draw [very thick] (9.6,8.0) to (9.6,8.4);
\draw [very thick] (9.6,8.4) to (9.6,8.8);
\draw [very thick] (9.6,8.8) to (10.,8.4);
\draw [very thick] (10,8.4) to (10,8.8);
\draw [very thick] (10,8.8) to (10.4,8.4);
\draw [very thick] (10.4,8.4) to (10.4,8.8);
\draw [very thick] (10.4,8.8) to (10.8,8.4);
\draw [very thick] (10.8,8.4) to (10.8,8.8);
\draw [very thick] (10.8,8.8) to (11.2,8.4);
\draw [very thick] (11.2,8.4) to (11.2,8.8);
\draw [very thick] (11.2,8.8) to (11.6,8.4); 
\draw [very thick]  (11.6,8.4) to  (11.6,8.8);
\draw [very thick]   (11.6,8.8) to  (12,8.4); 
\end{tikzpicture}}

\caption{\label{fig:a} $rp$ process path for 1.2 GK}
\end{figure} 

\begin{figure}
\center
\resizebox{10cm}{!}{

\begin{tikzpicture}

  \node (1) at (.4,0)  [draw,inner sep=5.8pt,label=left:$_{28}$Ni,label=below :28] {$ $  };
    \node (2) at (.4,.4) [draw,inner sep=5.8pt] {$ $};
    \node (3) at (.4,.8) [draw,inner sep=5.8pt,label=left:$_{30}$Zn] {$ $};
    \node (4) at (.4,1.2) [draw,inner sep=5.8pt] {$ $};
    \node (5) at (.8,.4) [draw,inner sep=5.8pt]{$ $};
    \node (6) at (.8,.8) [draw,inner sep=5.8pt]{$ $};
 \node (7) at (.8,1.2) [draw,inner sep=5.8pt]   {$ $ };
 \node (8) at (1.2,.4) [draw,inner sep=5.8pt]    {$ $};
 \node (9) at (1.2,.8) [draw,inner sep=5.8pt]    {$ $};
 \node (10) at (1.2,1.2) [draw,inner sep=5.8pt]   {$ $};
\node (11) at (1.2,1.6) [draw,inner sep=5.8pt,label=left:$_{32}$Ge]    {$ $};
 \node (12) at (1.6,.4) [draw,inner sep=5.8pt] {$ $};
 \node (13) at (1.6,.8) [draw,inner sep=5.8pt] {$ $};
 \node (14) at (1.6,1.2) [draw,inner sep=5.8pt] {$ $};
\node (15) at (1.6,1.6) [draw,inner sep=5.8pt]  {$ $};
 \node (16) at (2.0,1.2) [draw,inner sep=5.8pt]   {$ $};
\node (17) at (2.0,1.6) [fill,draw,inner sep=5.8pt]    {$ $};
 \node (18) at (2.0,2.0) [draw,inner sep=5.8pt]   {$ $};
\node (19) at (2.0,2.4) [draw,inner sep=5.8pt]    {$ $};
 \node (20) at (2.4,1.2) [draw,inner sep=5.8pt] {$ $};
 \node (21) at (.8,0) [draw,inner sep=5.8pt]  {$ $};
 \node (22) at (1.2,0) [draw,inner sep=5.8pt,label=below :30]   {$ $};
 \node (23) at (1.6,0) [draw,inner sep=5.8pt] {$ $};
 \node (24) at (2.0,0) [draw,inner sep=5.8pt,label=below :32]   {$ $};
 \node (25) at (2.0,.4) [draw,inner sep=5.8pt]   {$ $};
 \node (26) at (2.0,.8) [draw,inner sep=5.8pt]   {$ $};
\node (27) at (1.2,2.0) [draw,inner sep=5.8pt] {$ $};
\node (28) at (1.2,2.4) [draw,inner sep=5.8pt,label=left:$_{34}$Se] {$ $};
\node (29) at (1.6,2.0) [draw,inner sep=5.8pt] {$ $};
\node (30) at (1.6,2.4) [draw,inner sep=5.8pt] {$ $};
\node (31) at (2.4,1.6) [draw,inner sep=5.8pt] {$ $};
\node (32) at (2.4,2.0) [draw,inner sep=5.8pt] {$ $};
\node (33) at (2.4,2.4) [draw,inner sep=5.8pt] {$ $};
\node (34) at (2.4,2.8) [draw,inner sep=5.8pt] {$ $};
 \node (35) at (2.8,.8) [draw,inner sep=5.8pt,label=below :34] {$ $};
 \node (36) at (2.8,1.2) [draw,inner sep=5.8pt] {$ $};
 \node (37) at (2.8,1.6) [draw,inner sep=5.8pt] {$ $};
\node (38) at (2.8,2.0) [draw,inner sep=5.8pt] {$ $};
\node (39) at (2.8,2.4) [fill,draw,inner sep=5.8pt] {$ $};
\node (40) at (2.8,2.8) [draw,inner sep=5.8pt] {$ $};
\node (41) at (2.8,3.2) [draw,inner sep=5.8pt,label=left:$_{36}$Kr] {$ $};
\node (42) at (3.2,2.0) [draw,inner sep=5.8pt] {$ $};
\node (43) at (3.2,2.4) [draw,inner sep=5.8pt] {$ $};
\node (44) at (3.2,2.8) [draw,inner sep=5.8pt] {$ $};
\node (45) at (3.2,3.2) [draw,inner sep=5.8pt] {$ $};
\node (46) at (3.6,2.4) [draw,inner sep=5.8pt,label=below:36] {$ $ };
\node (47) at (3.6,2.8) [draw,inner sep=5.8pt] {$ $ };
\node (48) at (3.6,3.2) [fill,draw,inner sep=5.8pt] {$ $};
\node (49) at (3.6,3.6) [draw,inner sep=5.8pt] {$ $ };
\node (50) at (3.6,4.) [draw,inner sep=5.8pt,label=left:$_{38}$Sr] {$ $ };
\node (51) at (4.0,2.8) [draw,inner sep=5.8pt] {$ $ };
\node (52) at (4.0,3.2) [draw,inner sep=5.8pt] {$ $  };
\node (53) at (4.0,3.6) [draw,inner sep=5.8pt] {$ $ };
\node (54) at (4.0,4.) [draw,inner sep=5.8pt] {$   $ };
\node (55) at (4.0,4.4) [draw,inner sep=5.8pt] {$  $ };
\node (56) at (4.4,3.2) [draw,inner sep=5.8pt,label=below:38] {$ $ };
\node (57) at (4.4,3.6) [draw,inner sep=5.8pt] {$  $ };
\node (58) at (4.4,4.) [fill,draw,inner sep=5.8pt] {$ $ };
\node (59) at (4.4,4.4) [draw,inner sep=5.8pt] {$ $ };
\node (60) at (4.4,4.8) [draw,inner sep=5.8pt,label=left:$_{40}$Zr] {$ $ };
\node (61) at (4.8,3.6) [draw,inner sep=5.8pt] {$ $ };
\node (62) at (4.8,4.) [draw,inner sep=5.8pt] {$ $ };
\node (63) at (4.8,4.4) [draw,inner sep=5.8pt] {$ $ };
\node (64) at (4.8,4.8) [draw,inner sep=5.8pt] {$ $ };
\node (65) at (4.8,5.2) [draw,inner sep=5.8pt] {$ $ };
\node (66) at (5.2,4.) [draw,inner sep=5.8pt,label=below:40] {$ $ };
\node (67) at (5.2,4.4) [draw,inner sep=5.8pt] {$ $ };
\node (68) at (5.2,4.8) [fill,draw,inner sep=5.8pt] {$ $ };
\node (69) at (5.2,5.2) [draw,inner sep=5.8pt] {$ $ };
\node (70) at (5.2,5.6) [draw,inner sep=5.8pt,label=left:$_{42}$Mo] {$ $ };
\node (71) at (5.6,4.4) [draw,inner sep=5.8pt] {$ $ };
\node (72) at (5.6,4.8) [draw,inner sep=5.8pt] {$ $ };
\node (73) at (5.6,5.2) [draw,inner sep=5.8pt] {$ $ };
\node (74) at (5.6,5.6) [draw,inner sep=5.8pt] {$ $ };
\node (75) at (5.6,6.0) [draw,inner sep=5.8pt] {$ $ };
\node (76) at (6.0,4.8) [draw,inner sep=5.8pt,label=below:42] {$ $ };
\node (77) at (6.0,5.2) [draw,inner sep=5.8pt] {$ $ };
\node (78) at (6.0,5.6) [fill,draw,inner sep=5.8pt] {$ $ };
\node (79) at (6.0,6.0) [draw,inner sep=5.8pt] {$ $ };
\node (80) at (6.0,6.4) [draw,inner sep=5.8pt,label=left:$_{44}$Ru] {$ $ };
\node (81) at (6.4,5.2) [draw,inner sep=5.8pt] {$ $ };
\node (82) at (6.4,5.6) [draw,inner sep=5.8pt] {$ $ };
\node (83) at (6.4,6.0) [draw,inner sep=5.8pt] {$ $ };
\node (84) at (6.4,6.4) [draw,inner sep=5.8pt] {$ $ };
\node (85) at (6.4,6.8) [draw,inner sep=5.8pt] {$ $ };
\node (86) at (6.8,5.6) [draw,inner sep=5.8pt,label=below:44] {$ $ };
\node (87) at (6.8,6.0) [draw,inner sep=5.8pt] {$ $ };
\node (88) at (6.8,6.4) [fill,draw,inner sep=5.8pt] {$ $ };
\node (89) at (6.8,6.8) [draw,inner sep=5.8pt] {$ $ };
\node (90) at (6.8,7.2) [draw,inner sep=5.8pt,label=left:$_{46}$Pd] {$ $ };
\node (91) at (7.2,6.0) [draw,inner sep=5.8pt] {$ $ };
\node (92) at (7.2,6.4) [draw,inner sep=5.8pt] {$ $ };
\node (93) at (7.2,6.8) [draw,inner sep=5.8pt] {$ $ };
\node (94) at (7.2,7.2) [draw,inner sep=5.8pt] {$ $ };
\node (95) at (7.2,7.6) [draw,inner sep=5.8pt] {$ $ };
\node (96) at (7.6,6.4) [draw,inner sep=5.8pt,label=below:46] {$ $ };
\node (97) at (7.6,6.8) [draw,inner sep=5.8pt] {$ $ };
\node (98) at (7.6,7.2) [fill,draw,inner sep=5.8pt] {$ $ };
\node (99) at (7.6,7.6) [draw,inner sep=5.8pt] {$ $ };
\node (100) at (7.6,8.0) [draw,inner sep=5.8pt,label=left:$_{48}$Cd] {$ $ };
\node (101) at (8.0,6.8) [draw,inner sep=5.8pt] {$ $ };
\node (102) at (8.0,7.2) [draw,inner sep=5.8pt] {$ $ };
\node (103) at (8.0,7.6) [draw,inner sep=5.8pt] {$ $ };
\node (104) at (8.0,8.0) [draw,inner sep=5.8pt] {$ $ };
\node (105) at (8.0,8.4) [draw,inner sep=5.8pt] {$ $ };
\node (106) at (8.4,7.2) [draw,inner sep=5.8pt,label=below:48] {$ $ };
\node (107) at (8.4,7.6) [draw,inner sep=5.8pt] {$ $ };
\node (108) at (8.4,8.0) [fill,draw,inner sep=5.8pt] {$ $ };
\node (109) at (8.4,8.4) [draw,inner sep=5.8pt] {$ $ };
\node (110) at (8.4,8.8) [draw,inner sep=5.8pt,label=left:$_{50}$Sn] {$ $ };
\node (111) at (8.8,7.6) [draw,inner sep=5.8pt] {$ $ };
\node (112) at (8.8,8.0) [draw,inner sep=5.8pt] {$ $ };
\node (113) at (8.8,8.4) [draw,inner sep=5.8pt] {$ $ };
\node (114) at (8.8,8.8) [draw,inner sep=5.8pt] {$ $ };
\node (115) at (8.8,9.2) [draw,inner sep=5.8pt] {$ $ };
\node (116) at (9.2,8.0) [draw,inner sep=5.8pt,label=below:50] {$ $ };
\node (117) at (9.2,8.4) [draw,inner sep=5.8pt] {$ $ };
\node (118) at (9.2,8.8) [fill,draw,inner sep=5.8pt] {$ $ };
\node (119) at (9.2,9.2) [draw,inner sep=5.8pt] {$ $ };
\node (120) at (9.2,9.6) [draw,inner sep=5.8pt,label=left:$_{52}$Te] {$ $ };
\node (121) at (9.6,8.) [draw,inner sep=5.8pt] {$ $ };
\node (122) at (9.6,8.4) [draw,inner sep=5.8pt] {$ $ };
\node (123) at (9.6,8.8) [draw,inner sep=5.8pt] {$ $ };
\node (124) at (9.6,9.2) [draw,inner sep=5.8pt] {$ $ };
\node (125) at (9.6,9.6) [draw,inner sep=5.8pt] {$ $ };
\node (126) at (9.6,10.0) [draw,inner sep=5.8pt] {$ $ };
\node (127) at (10.0,8.0) [draw,inner sep=5.8pt,label=below:52] {$ $ };
\node (128) at (10.0,8.4) [draw,inner sep=5.8pt] {$ $ };
\node (129) at (10.0,8.8) [draw,inner sep=5.8pt] {$ $ };
\node (130) at (10.0,9.2) [draw,inner sep=5.8pt] {$ $ };
\node (131) at (10.0,9.6) [fill,draw,inner sep=5.8pt] {$ $ };
\node (132) at (10.0,10.0) [draw,inner sep=5.8pt] {$ $ };
\node (133) at (10.4,8.4) [draw,inner sep=5.8pt] {$ $ };
\node (134) at (10.4,8.8) [draw,inner sep=5.8pt] {$ $ };
\node (135) at (10.4,9.2) [draw,inner sep=5.8pt] {$ $ };
\node (136) at (10.4,9.6) [draw,inner sep=5.8pt] {$ $ };
\node (137) at (10.4,10.0) [draw,inner sep=5.8pt] {$ $ };
\node (138) at (10.8,8.4) [draw,inner sep=5.8pt] {$ $ };
\node (139) at (10.8,8.8) [draw,inner sep=5.8pt] {$ $ };
\node (140) at (10.8,9.2) [draw,inner sep=5.8pt] {$ $ };
\node (141) at (10.8,9.6) [draw,inner sep=5.8pt] {$ $ };
\node (142) at (10.8,10.0) [draw,inner sep=5.8pt] {$ $ };
\node (143) at (11.2,8.4) [draw,inner sep=5.8pt] {$ $ };
\node (144) at (11.2,8.8) [draw,inner sep=5.8pt] {$ $ };
\node (145) at (11.2,9.2) [draw,inner sep=5.8pt] {$ $ };
\node (146) at (11.2,9.6) [draw,inner sep=5.8pt] {$ $ };
\node (147) at (11.6,8.4) [draw,inner sep=5.8pt] {$ $ };
\node (148) at (11.6,8.8) [draw,inner sep=5.8pt] {$ $ };
\node (149) at (11.6,9.2) [draw,inner sep=5.8pt] {$ $ };
\node (150) at (12.0,8.4) [draw,inner sep=5.8pt] {$ $ };
\node (151) at (2.4,.8) [draw,inner sep=5.8pt] {$ $ };
\node (152) at (12.,8.8) [draw,inner sep=5.8pt] {$ $ };
\node (153) at (12.,9.2) [draw,inner sep=5.8pt] {$ $ };
\node (154) at (11.6,9.6) [draw,inner sep=5.8pt] {$ $ };
\node (155) at (12.,9.6) [draw,inner sep=5.8pt] {$ $ };
\node (156) at (12.,10.0) [draw,inner sep=5.8pt] {$ $ };
\node (157) at (11.6,10.0) [draw,inner sep=5.8pt] {$ $ };
\node (158) at (11.2,10.0) [draw,inner sep=5.8pt] {$ $ };

\draw [very thick]  (.4,0) to (.4,.4); 
\draw [very thick]  (.4,.4) to (.4,.8); 
\draw [very thick]  (.4,.8) to (0.8,.4); 
\draw [very thick]  (.8,.4) to (.8,.8); 
\draw [very thick]  (.8,.8) to (1.2,.4); 
\draw [very thick]  (1.2,.4) to (1.2,.8); 
\draw [very thick]  (1.2,.8) to (1.2,1.2); 
\draw [very thick]  (1.2,1.2) to (1.2,1.6); 
\draw [very thick]  (1.2,1.6) to (1.6,1.2);
\draw [very thick]  (1.6,1.2) to (1.6,1.6);
\draw [very thick]  (1.6,1.6) to (2.,1.2); 
\draw [very thick]  (2.,1.2) to (2.,1.6);
\draw [very thick]  (2.,1.6) to (2.,2.);
\draw [very thick]  (2.,1.6) to (2.4,1.2);
\draw [very thick]  (2.4,1.2) to (2.4,2.0);
\draw [very thick]  (2.,2.) to (2.,2.4);
\draw [very thick]  (2.,2.4) to (2.4,2.);
\draw [very thick]  (2.4,2.) to (2.4,2.4);
\draw [very thick]  (2.4,2.4) to (2.8,2.0);
\draw [very thick]  (2.8,2.) to (2.8,2.4);
\draw [very thick]  (3.2,2.8) to (3.2,3.2);
\draw [very thick]  (3.2,3.2) to (3.6,2.8);
\draw [very thick]  (3.6,2.8) to (3.6,3.2);
\draw [thick,gray]  (3.6,3.2) to (3.6,3.6);
\draw [thick,gray]  (3.6,3.6) to (3.6,4.);
\draw [thick,gray]  (3.6,4) to (4,3.6);
\draw [very thick]  (4,3.6) to (4,4);
\draw [very thick]  (4,4.) to (4.4,3.6);
\draw [very thick] (4.4,3.6) to (4.4,4);
\draw [thick,gray] (4.4,4) to (4.4,4.4);
\draw [thick,gray] (4.4,4.4) to (4.4,4.8);
\draw [thick,gray] (4.4,4.8) to (4.8,4.4);
\draw [very thick] (4.8,4.4) to (4.8,4.8);
\draw [very thick] (4.8,4.8) to (5.2,4.4);
\draw [very thick] (5.2,4.4) to (5.2,4.8);
\draw [very thick] (3.6,3.2) to (4.0,2.8);
\draw [very thick] (4,2.8) to (4,3.2);
\draw [very thick] (4,3.2) to (4,3.6);
\draw [very thick] (4,3.6) to (4,4.);
\draw [very thick] (4.4,4) to (4.8,3.6);
\draw [very thick] (4.8,3.6) to (4.8,4);
\draw [very thick] (4.8,4) to (4.8,4.4);
\draw [very thick] (4.8,4.4) to (4.8,4.8);
\draw [very thick] (2.8,2.4) to (3.2,2.);
\draw [very thick] (3.2,2.) to (3.2,2.4);
\draw [very thick] (3.2,2.4) to (3.2,2.8);
\draw [very thick] (5.2,4.8) to (5.6,4.4);

\draw [very thick] (5.6,4.4) to (5.6,4.8);
\draw [very thick] (5.6,4.8) to (5.6,5.2);
\draw [very thick] (5.6,5.2) to (5.6,5.6);
\draw [very thick] (5.6,5.2) to (6.,4.8);
\draw [very thick] (5.6,5.6) to (6,5.2);
\draw [gray, thick] (6.,4.8) to (6.,5.2);
\draw [very thick] (6.,5.2) to (6.,5.6);
\draw [very thick] (6.,5.6) to (6.4,5.2);
\draw [very thick] (6.4,5.2) to (6.4,5.6);
\draw [very thick] (6.4,5.6) to (6.4,6);
\draw [very thick] (6.4,6) to (6.4,6.4);
\draw [very thick] (6.4,6) to (6.8,5.6);
\draw [thick,gray] (6.4,6.4) to (6.8,6.);
\draw [gray, thick] (6.8,5.6) to (6.8,6);
\draw [very thick] (6.8,6) to (6.8,6.4);
\draw [very thick] (6.8,6.4) to (7.2,6);
\draw [very thick] (7.2,6) to (7.2,6.4);
\draw [very thick] (7.2,6.4) to(7.2,6.8);
\draw [very thick](7.2,6.8) to (7.2,7.2);
\draw [very thick](7.2,6.8) to (7.6,6.4);
\draw [very thick] (7.2,7.2) to (7.6,6.8);
\draw [very thick] (7.6,6.8) to (7.6,7.2);
\draw [very thick] (7.6,7.2) to (8,6.8);
\draw [very thick] (8,6.8) to (8,7.2);
\draw [very thick] (8,7.2) to (8,7.6);
\draw [very thick] (8,7.6) to (8,8.);
\draw [very thick] (8,7.6) to (8.4,7.2);
\draw [very thick] (8.0,8.0) to (8.4,7.6);
\draw [very thick] (8.4,7.2) to (8.4,7.6);
\draw [very thick] (8.4,7.6) to (8.4,8.0);
\draw [very thick](8.4,8.0) to (8.8,7.6);
\draw [very thick] (8.8,7.6) to (8.8,8.0);
\draw [very thick] (8.8,8.0) to (8.8,8.4);
\draw [very thick] (8.8,8.4) to (9.2,8.0);
\draw [very thick] (9.2,8.0) to (9.2,8.4);
\draw [very thick] (9.2,8.4) to (9.6,8.0);
\draw [very thick] (9.6,8.0) to (9.6,8.4);
\draw [very thick] (9.6,8.4) to (9.6,8.8);
\draw [very thick] (9.6,8.8) to (10.,8.4);
\draw [very thick] (10,8.4) to (10,8.8);
\draw [very thick] (10,8.8) to (10.4,8.4);
\draw [very thick] (10.4,8.4) to (10.4,8.8);
\draw [very thick] (10.4,8.8) to (10.8,8.4);
\draw [very thick] (10.8,8.4) to (10.8,8.8);
\draw [very thick] (10.8,8.8) to (11.2,8.4);
\draw [very thick] (11.2,8.4) to (11.2,8.8);
\draw [very thick] (11.2,8.8) to (11.6,8.4); 
\draw [very thick]  (11.6,8.4) to  (11.6,8.8);
\draw [very thick]   (11.6,8.8) to  (12,8.4); 
\end{tikzpicture}}

\caption{\label{fig:b} $rp$ process path for 1.5 GK}
\end{figure} 

The $rp$ process paths are shown in Fig. 1 and Fig. 2 
for temperatures 1.2 and 1.5 GK, respectively, using a constant proton flux density 
of $10^{6} gram/cm^{3}$,  a proton fraction of 0.7 and 100 seconds burst 
duration. Here, 
black lines indicate the path along which the major portion of
 the total flux flow whereas gray lines indicate the minor paths. The filled
 boxes in figures indicate the waiting points. As evident from these figures,
$rp$ process paths depend on the temperature of the environment to some extent. For example, 
 at T= 1.2 GK, major portions of the $rp$ process flux at the waiting point nucleus $^{64}$Ge   
  convert to  $^{66}$Se by 
two-proton capture process. Less than 1\% of the total flux flows through the
 $\beta$-decay channel and follow
  the paths showed by  gray lines in Fig. 1. In contrast, 
at temperature T= 1.5 GK (in Fig. 2), the probability of two-proton capture of $^{64}$Ge gets reduced and
the $rp$ process path bifurcates from the waiting point nucleus  almost in an equal proportion. This suggests that,
at 1.2 GK, proton capture by  $^{64}$Ge dominates over its inverse process,
i.e. photodisintegration. As the temperature increases, contributions of 
photodisintegration process
increases and therefore a large fraction of the  total abundance chooses another path 
through more stable nuclei such as $^{65}$Ge, $^{66}$As, etc. Similarly, near other waiting points ($\it{viz.}$
 $^{72}$Kr, $^{76}$Sr etc.) the abundance flow pattern changes with changing temperature. It is also evident from above 
the figures
that, beyond mass 80, temperature change can hardly affect the scenario.

  As shown in
Fig. 1 and Fig. 2, above A=100 region, the $rp$ process
continues through proton capture by In isotopes and $\beta$-decay of Sn 
isotopes. Here, $^{100}$In
captures a proton to form $^{101}$Sn which 
completely decays to $^{101}$In as $^{102}$Sb is proton unbound. In turn, $^{101}$In undergoes proton 
capture and further exhibits $\beta$-decay. The process
 of proton capture followed by consecutive
$\beta$-decay continues and relative abundances of
nuclei decrease as one proceeds towards higher
mass region. Ultimately, less than 0.001\% of
the total flux can reach $^{106}$In, according to our
present calculation.

 According to the calculations of Schatz {\em et al.}\cite{schatz}, 
a significant portion of the $^{105}$Sn captures a proton to form $^{106}$Sb, as $^{106}$Sb has 
a sufficient positive proton separation energy (0.59 MeV from FRDM\cite{Mol} calculation). 
Another proton capture leads to $^{107}$Te, which instantly undergoes 
$\alpha$-decay to $^{103}$Sn. Thus, the $rp$ process ends in
the SnSbTe cycle.

In contrast, our calculation does not go through the SnSbTe cycle.
It is clear that the SnSbTe cycle may not occur under two scenarios. 
Firstly, if $^{106}$Sb isotope be very loosely bound (proton 
separation energy 0.119 MeV according to the mass formula\cite{mass}), 
any $^{106}$Sb that is 
formed by a proton capture instantly reverts back to $^{105}$Sn. Second 
scenario occurs when
the proton capture rates are too small for $^{105}$Sn to initiate a proton capture process. 

In a previous work\cite{epja}, we have shown that a small 
fluctuation in the mass values of waiting point nuclei in mass 60-80 
region may affect the effective half life and thus, proper knowledge of 
ground state binding energy is necessary to understand the 
bridging phenomena of a waiting point nucleus below mass 80 
region. However, as we move towards the higher mass region, we find
that small variations in binding energy do not affect the $rp$ process
path significantly. 
Taking the rms error into account for the
 proton separation energy of $^{106}$Sb $(0.119 \pm 0.376)$MeV, we have  
checked whether the proton capture on $^{105}$Sn can dominate over 
its inverse process.  We find that the fraction of the initial 
flux entering into the SnSbTe-cycle is negligibly small. We have 
repeated our calculation with the proton separation 
energy  of $^{106}$Sb isotope from a recent experiment by Elomma 
{\em et al.}\cite{elomma}, {\em viz.} 0.428(0.008)MeV.
We find that our results remain almost invariant. Repeating
the entire calculation with ground state binding energies from FRDM\cite{Mol}
 calculation or the  
Duflo-Zuker\cite{DZ} mass formula do not alter this conclusion. 
Hence, we see that for a reasonable variation of mass values
the $rp$ process fails to enter into the SnSbTe-cycle 
for an  X-ray burst of 100 seconds duration. If we consider an X-ray 
burst having duration greater than 150 seconds, a very small fraction of 
total flux,  less than 0.001\% of the original, enters the cycle. 

 \begin{figure}[htb]
\center
\resizebox{6.5cm}{5cm}{\includegraphics{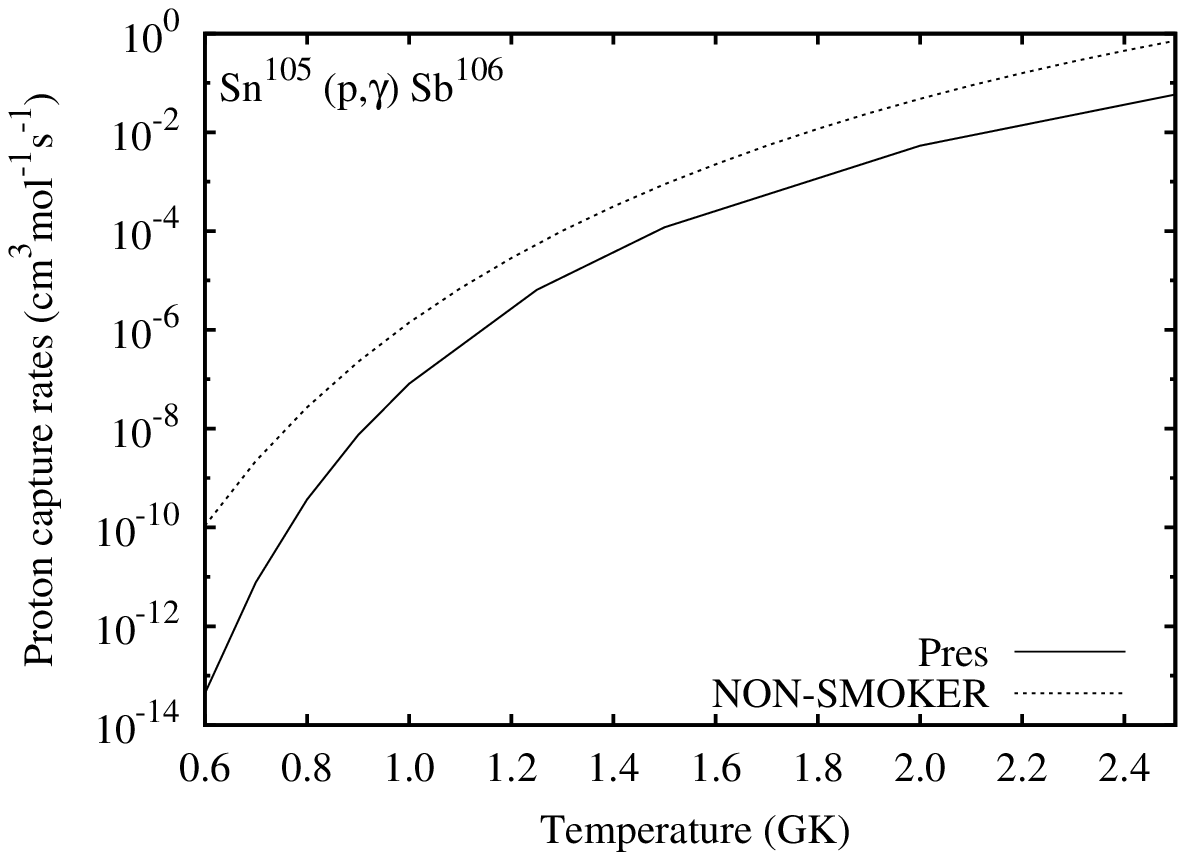}}\hskip -0.4cm 
\resizebox{6.5cm}{5cm}{\includegraphics{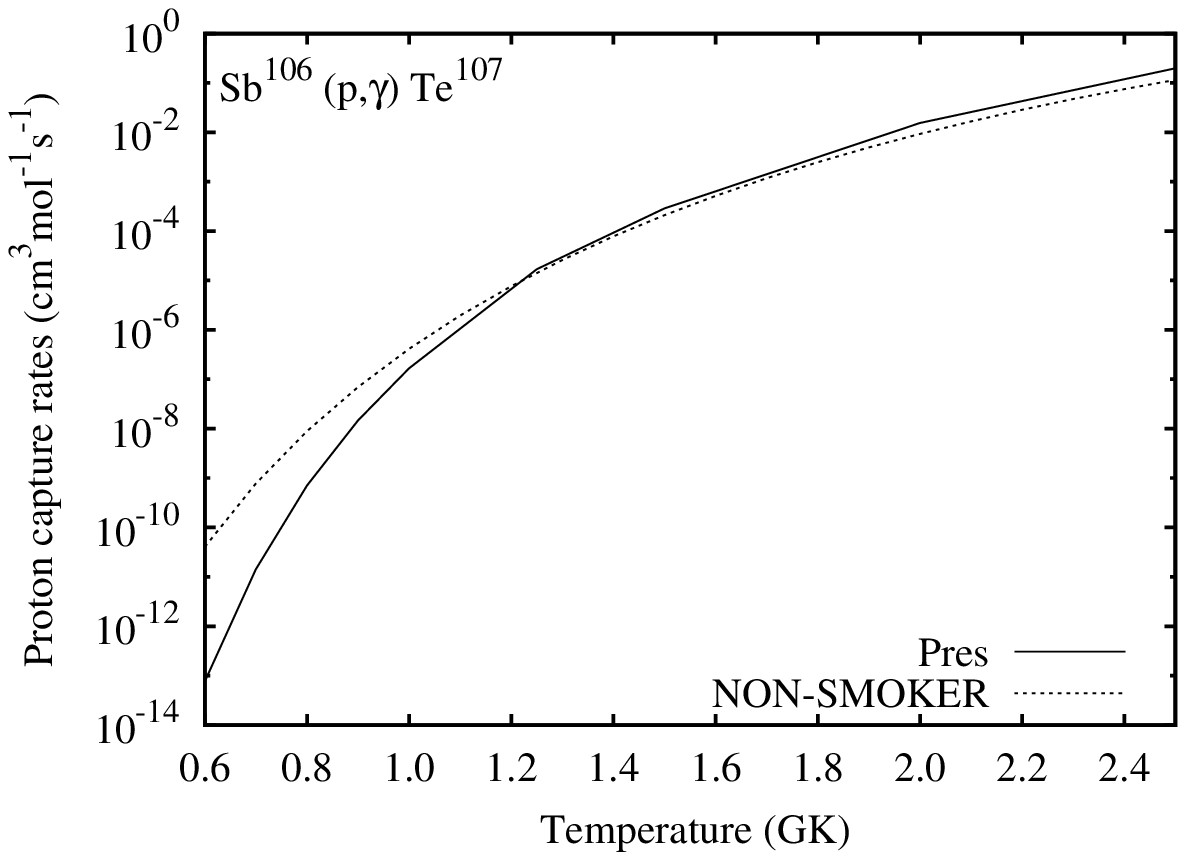}} 
\caption{\label{fig:c} Proton capture rates ($cm^{3}mol^{-1}sec^{-1}$) vs temperature(GK). See text for details.} 
\end{figure}

Schatz {\em et al.}\cite{schatz} have used capture rates different from ours.
In Fig. 3, we have plotted proton capture rates with temperature for the isotopes $^{105}$Sn
 and $^{106}$Sb respectively. The different values
have been indicated as follows: Pres - Present work and
NON-SMOKER - NON-SMOKER\cite{ns,ns1} results used by Schatz {\em et al.}\cite{schatz}
 %
In case of $^{105}$Sn nuclei, it is evident from Fig. 3 that proton capture rates from our calculation
are approximately $10^2$ times smaller than the rates from NON-SMOKER calculation. It seems, the difference 
is mainly due to the fact that they have used the form of the interaction
 from Jeukenne {\em et al.}\cite{jlm}
 which is different from our case. For better understanding, we have
repeated the entire calculation using reaction rates from Hauser-Feshbach code NON-SMOKER\cite{ns,ns1}. In this case, 
considerable fraction of the initial flux ($\sim$ 0.1\%) enters into the SnSbTe cycle. 
 Thus, we may conclude that 
the  proton capture rates, and not the mass values, determine 
the endpoint of $rp$ process. In our opinion, our approach in fixing the 
parameters in the reaction calculation by fitting known reaction 
rates, and extrapolating the calculation to unknown reactions, 
may be relied upon to predict the endpoint correctly.

There are many X-ray burster models in literature with various 
density-temperature profiles. Our present goal is to study how different 
temperature-density profiles can affect the relative abundances of the nuclei produced by
 $rp$ process and the endpoint. In Fig. 4, we have plotted 
the relative abundances of elements versus mass number at 100th second of the 
burst in various models, as described below. 

\begin{figure}[htb]
\center
\resizebox{6.5cm}{5cm}{\includegraphics{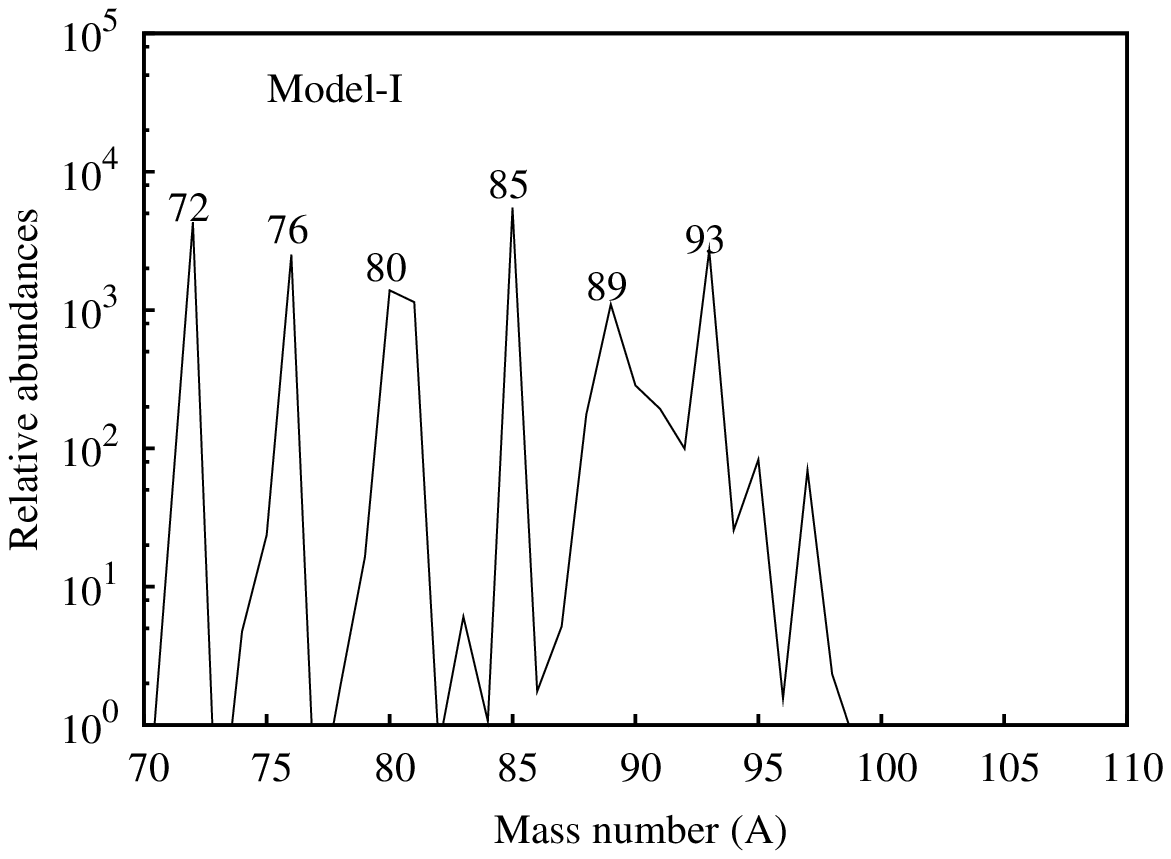}}\hskip -0.4cm 
\resizebox{6.5cm}{5cm}{\includegraphics{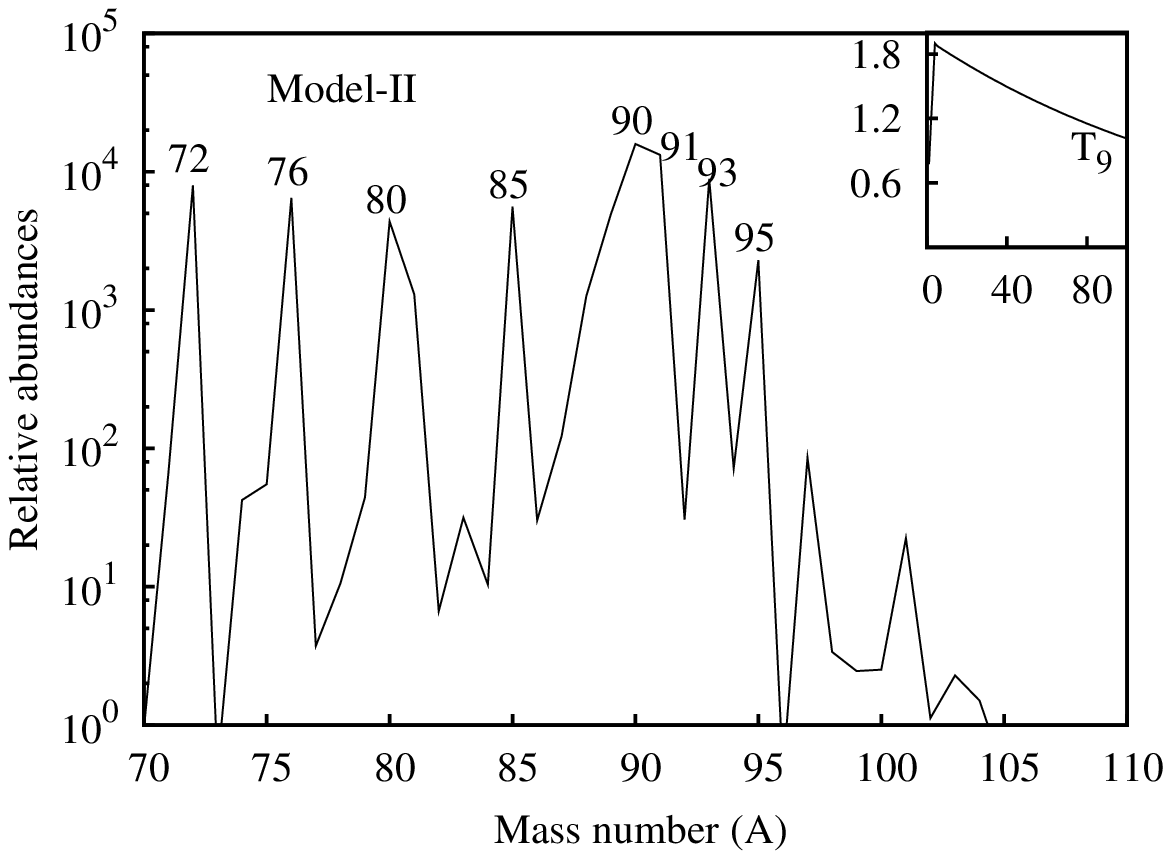}} 
\resizebox{6.5cm}{5cm}{\includegraphics{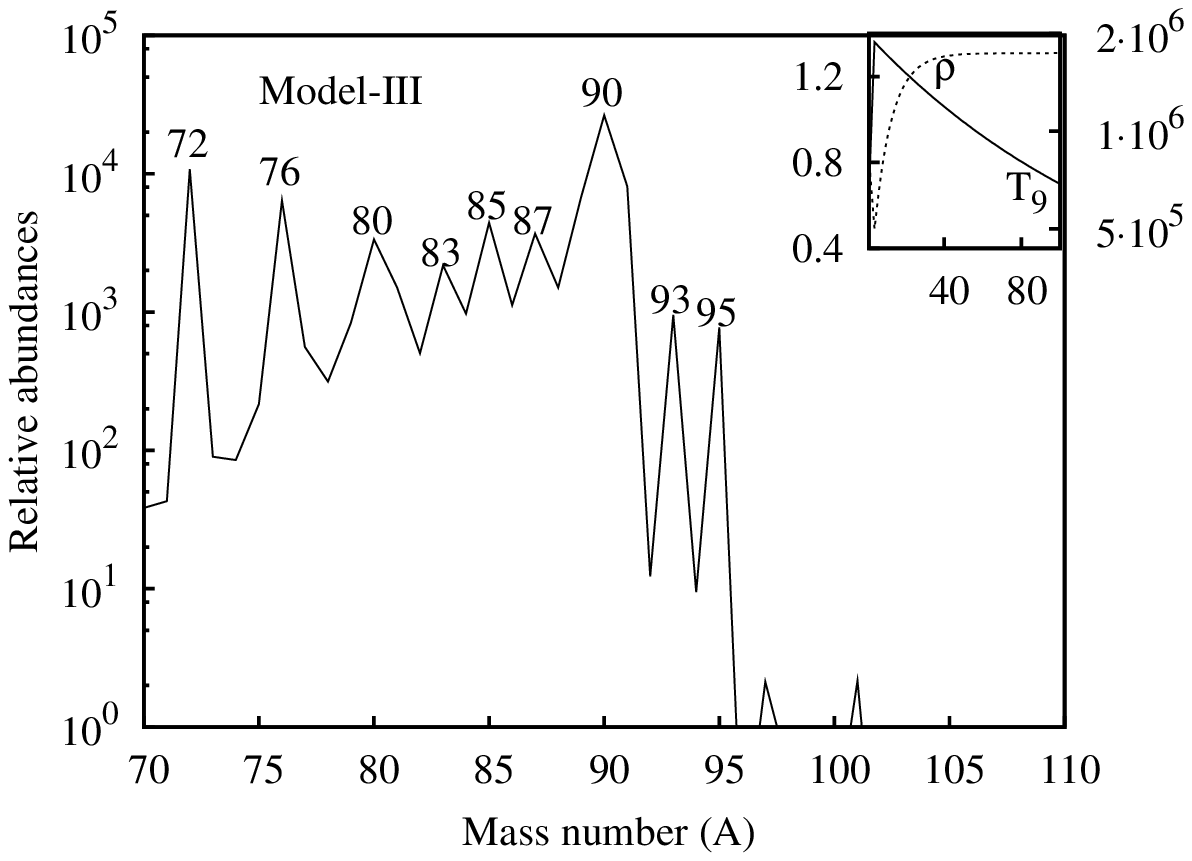}}\hskip -0.35cm 
\resizebox{6.5cm}{5cm}{\includegraphics{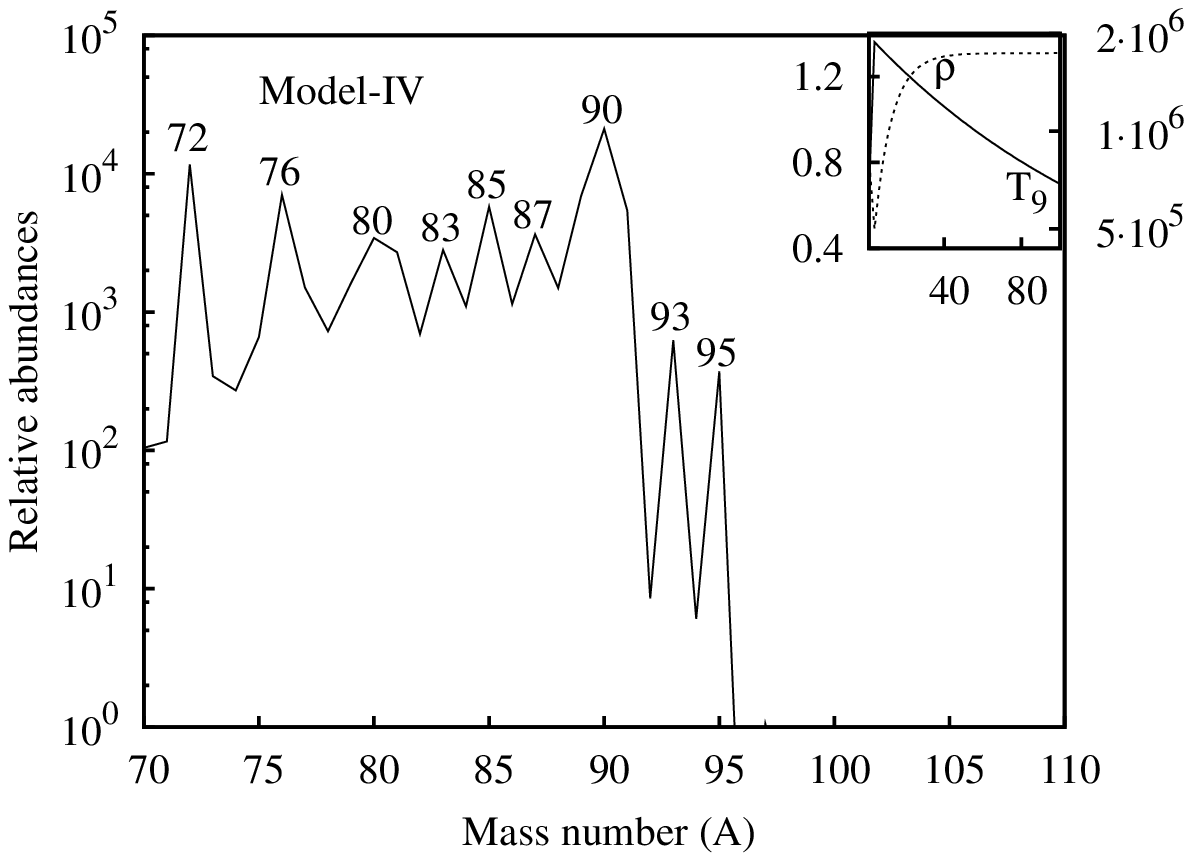}}
\caption{\label{fig:d} Relative abundances vs mass number for different density-temperature 
profiles of various X-ray burster after 100 seconds. (Insets) Model-II: Curve for temperature in GK 
vs time in seconds of the X-ray burst from Schatz {\em et al.}\cite{schatz}; Model-III and IV: Temperature in GK (left)  
and density $\rho$ in $gram/cm^3$ (right)  with 
time in seconds from Illiadis\cite{iliadis}. See text for details.} 


\end{figure}

In the first case (Model-I), we assume a constant density-temperature
framework with temperature $T = 1.5$ GK, 
density $\rho=10^{6} gram/cm^{3}$ and proton fraction value 0.7.  
Model-II\cite{schatz} describes a situation where ignition takes place at a
constant density of $\rho= 1.1 \times 10^{6} gram/cm^{3}$ and the 
burst reaches a peak temperature $T_{peak}$ = 1.9 GK 
after 4 seconds while the cooling phase lasts for approximately 200 seconds. 
In another example, taken from the book by Illiadis\cite{iliadis} (Model-III), 
nuclear burning starts with temperature and density values of $T = 0.4$ GK 
and $\rho=10^{6} gram/cm^{3}$, respectively. After 4 seconds, a maximum temperature 
of $T_{peak}$ = 1.36 GK and a minimum density of
$\rho_{peak} = 5 \times 10^{5} gram/cm^{3}$ are achieved. After 
100 seconds, the temperature drops to $T = 0.7$ GK and the density 
increases to $\rho = 1.4 \times 10^{6} gram/cm^{3}$. For all the above
models, 
it is assumed that the X-ray burst environment is sufficiently proton-rich 
to maintain the number 0.7 as the constant proton fraction. As a significant 
amount of proton flux is consumed during the thermonuclear reaction,
another situation may arise, where the proton fraction decreases gradually 
with time. Such a situation is given in Model-IV   where the proton fraction 
decreases to 0.16 after 100 seconds with temperature-density profile same as
that of the Model-III. It is evident from Fig. 4 that abundance peaks 
in all cases are obtained at mass values of 72, 76 and 80,  as a result of
the existence of waiting point nuclei $^{72}$Kr, $^{76}$Sr and $^{80}$Zr, 
respectively. Other peaks (for example, peaks at mass 85, 93 and 95  for 
isotopes $^{85}$Mo, $^{93}$Pd, $^{95}$Cd, respectively) suggest that 
the $rp$ process flux gets accumulated at these points due to very small 
positive or negative proton separation energies of those isotopes.  
We have considered the mass number in each case for which the 
flux amount drops below 1\% of the initial flux. It is evident from Fig. 4 
that for Model-I, A $\approx$ 93 is the region above which 
the $rp$ process flux fall below the range of our interest, whereas 
for Model-II, the region is around A $\approx$ 95. 
Other abundance peaks are observed at masses $A=$97 and 101 for the isotopes 
$^{97}$Cd and $^{101}$Sn respectively, though, 
the fractions of the total flux accumulated at those isotopes are less than 0.1\% of the initial value.
In case of Models III and IV, the locations to be studied
 are around  A $\approx$  93 and 91 respectively. 
From above observations,
it can be concluded that, for various density-temperature profiles, 
the $rp$ process flux falls below a significant amount 
near mass 90-95. 
The observations from Fig. 4  suggest that the end points of the $rp$ process has a rather
weak dependence on different X-ray burster models.

\section{Summary}
The location above which the $rp$ process flux falls below an 
insignificant amount is calculated using the microscopic optical model
utilizing the densities from the RMF approach and with a new 
phenomenological mass formula. Present result is compared with 
result obtained from another existing work\cite{schatz} and the reason for these 
indifferences between the results are discussed.
Our results do not significantly depend on the mass models.
For different X-ray burster models, endpoints are calculated.     

\section*{Acknowledgments}
This work has been carried out with financial assistance of the UGC sponsored
DRS Programme of the Department of Physics of the University of Calcutta.
Chirashree Lahiri acknowledges the grant of a fellowship awarded by the UGC.


\begin{thebibliography}{99}
\bibitem{mass} G. Gangopadhyay, {\it Int. J. Mod. Phys.} E  {\bf 20} (2011) 179.
\bibitem{schatz} H. Schatz {\em et al.} \PRL {\bf 86} (2001) 3471.
\bibitem{Mol}P. M\"{o}ller, J. R. Nix, and K. L. Kratz,  \ADNDT {\bf 66} (1997) 131.
\bibitem{ns} T. Rauscher and F.-K. Thielmann, \ADNDT {\bf 75} (2000) 1.
\bibitem{ns1} T. Rauscher and F.-K. Thielmann, \ADNDT {\bf 76} (2001) 47.
\bibitem{prc} C. Lahiri and G. Gangopadhyay, {\it Phys. Rev.} C {\bf 84} (2011) 057601.
\bibitem{gold}B. G. Todd-Rutel and J. Piekarewicz, \PRL {\bf 95} (2005) 122501.
\bibitem{1}M. Bhattacharya and G. Gangopadhyay, \PR C {\bf 77} (2008) 047302.
\bibitem{2} M. Bhattacharya, G. Gangopadhyay and S. Roy, \PR C {\bf 85} (2012) 034312.
\bibitem{ddm3y1} A.M. Kobos, B.A. Brown, R. Lindsay, G.R. and Satchler, \NP A {\bf 425} (1984) 205.
\bibitem{ddm3y2}A.K. Chaudhuri,  \NP A {\bf 449} (1986) 243.
\bibitem{ddm3y3}A.K. Chaudhuri,  \NP A {\bf 459} (1986) 417.
\bibitem{epja} C. Lahiri and G. Gangopadhyay, \EPJ   A {\bf 47} (2011) 87.
\bibitem{talys}A. J. Koning, S. Hilaire, and M. Duijvestijn, Proceedings of
the International Conference on Nuclear Data for Science
and Technology, April 22–27, 2007, Nice, France, edited by
O. Bersillon, F. Gunsing, E. Bauge, R. Jacqmin, and S. Leray
(EDP Sciences, Paris, 2008), pp. 211–214.
\bibitem{ijmpe} C. Lahiri and G. Gangopadhyay, {\it Int. J. Mod. Phys.} E 
{\bf 20} (2011) 2417. 
\bibitem{nrich}C. Lahiri and G. Gangopadhyay, to appear in 
{\it Int. J. Mod. Phys.} E (2012).
\bibitem{iliadis} C. Illiadis, {\it Nuclear Physics of the Stars} 
(Wiley-VCH Verlag GmbH, Weinheim, 2007).
\bibitem{ptable} T. Rauscher and F-K Thielemann, \ADNDT {\bf 74} (2000) 1.
\bibitem{audi14}G. Audi, O. Bersillon, J. Blachot  and A.H. Wapstra,
{\it Nucl. Phys.} A {\bf 729}  (2003) 3.
\bibitem{lopez} M.J. L\'{o}pez Jim\'{e}nez {\em et al.}, {\it Phys. Rev.} C {\bf 66}
(2002) 025803.
\bibitem{elomma} V.- V. Elomaa {\em et al.} \PRL {\bf 102} (2009) 252501.
\bibitem{DZ}J. Duflo and A.P. Zuker,  {\it Phys. Rev.} C {\bf 52} (1995) R23.
\bibitem{jlm}J. Jeukenne, A. Lejeune, and C. Mahaux,  \PR C {\bf 16} (1977) 80.

\end{thebibliography}
\end{document}